# Assessment of Density Functional Methods for Exciton Binding Energies and Related Optoelectronic Properties

Jui-Che Lee,[†] Jeng-Da Chai*[‡] and Shiang-Tai Lin*[†]

[†]Department of Chemical Engineering, National Taiwan University, Taipei 10617, Taiwan.
E-mail: stlin@ntu.edu.tw

[‡]Department of Physics, Center for Theoretical Sciences, and Center for Quantum Science and Engineering, National Taiwan University, Taipei 10617, Taiwan.
E-mail: jdchai@phys.ntu.edu.tw

**ABSTRACT**

The exciton binding energy, the energy required to dissociate an excited electron-hole pair into free charge carriers, is one of the key factors to the optoelectronic performance of organic materials. However, it remains unclear whether modern quantum-mechanical calculations, mostly based on Kohn-Sham density functional theory (KS-DFT) and time-dependent density functional theory (TDDFT), are reliably accurate for exciton binding energies. In this study, the exciton binding energies and related optoelectronic properties (e.g., the ionization potentials, electron affinities, fundamental gaps, and optical gaps) of 121 small- to medium-sized molecules are calculated using KS-DFT and TDDFT with various density functionals. Our KS-DFT and TDDFT results are compared with those calculated using highly accurate CCSD and EOM-CCSD methods, respectively. The ωB97, ωB97X, and ωB97X-D functionals are shown to generally outperform (with a mean absolute error of 0.36 eV) other functionals for the properties investigated.

# I. Introduction

An exciton, a bound electron-hole pair, can be generated when light is absorbed in a photoactive material. The electron-hole pair may be located at the same molecular unit (e.g., a Frenkel exciton) or different molecular units (e.g., an intermolecular charge-transfer exciton).[1] The Coulomb interaction that stabilizes the exciton with respect to free electron and hole is known as the exciton binding energy.[2] Materials with small exciton binding energies usually exhibit high charge separation efficiency,[3] which are often desirable for photovoltaic applications, whereas the opposite may be favorable for light-emitting devices[1]. Covalently bounded inorganic semiconductors have delocalized charge carriers and broad valence and conduction bands with exciton binding energies being as small as a few millielectron volts (meV). In contrast, the electronic properties of organic semiconductors are dominated by the localized charge on individual molecules and the large polarizabilities.[4, 5] As a result, the exciton binding energies of organic semiconductors can be as large as 0.1-1 eV. Therefore, it is important to develop a comprehensive understanding of the relevant parameters controlling the exciton binding energies of organic materials for the design of ideal photoactive material.

While a direct measurement of the exciton binding energy may be challenging, the exciton binding energy can be obtained from the difference between the fundamental and optical gaps, each of which can be measured directly.[4, 6-9] On the other hand, the fundamental and optical gaps can also be calculated using quantum-mechanical methods. Nayak et al.

calculated the exciton binding energies of small organic conjugated molecules in vacuum and in thin films[10, 11] using Kohn-Sham density functional theory (KS-DFT)[12, 13] and time-dependent density functional theory (TDDFT)[14, 15] with the B3LYP functional.[16, 17] However, the overall accuracy of KS-DFT and TDDFT with conventional density functionals is not comprehensively examined on exciton binding energies.

While intermolecular charge-transfer excitons are also important, in this work, we only focus on the Frenkel excitions. Specifically, we examine the accuracy of exciton binding energies and related optoelectronic properties on a diverse range of molecules using KS-DFT and TDDFT with several widely used density functionals. The rest of this paper is organized as follows. In Section II, we describe our test sets and computational details. The exciton binding energies and related optoelectronic properties obtained from density functional methods are compared with those obtained from high-level *ab initio* methods in Section III. Our conclusions are given in Section IV.

**II. Test sets and computational details**

To evaluate the performance of the functionals on exciton binding energies and related optoelectronic properties (e.g., vertical ionization potentials, vertical electron affinities, fundamental gaps, optical gaps, and exciton binding energies), we collect a test set, which consists of experimental vertical ionization potentials of 121 small- to medium-sized molecules in the experimental geometries. The geometries and reference values are taken from the IP131 database.[18] The nine molecules (C4H5N, C6H6, CF3CN, CH2ClCH2CH3, CH3CH(CH3)CH3, CH3CONH2, CH3SOCH3, N(CH3)3, SF6) were excluded because of the excessive computational resources needed. The coupled-cluster theory with iterative singles and doubles (CCSD)[19-21] and equation-of-motion CCSD (EOM-CCSD),[22-30] are employed as the benchmarks for the electronic and optical properties, respectively.

We examine the vertical ionization potentials (IP), vertical electron affinities (EA), fundamental gaps ($E_g$), optical gaps ($E_{opt}$), and exciton binding energies ($E_b$) of 121 molecules using various density functional methods. The IP, EA, and $E_g$ calculations follows the procedure detailed in our previous work.[31] For the KS-DFT and TDDFT calculations, we adopt popular density functionals, involving the local-density-approximation (LDA[32, 33]) functional, a generalized-gradient-approximation (GGA) functional (PBE[34, 35]), a meta-GGA functional (M06-L[36]), a global hybrid GGA functional (B3LYP[16, 17]), three long-range corrected (LC) hybrid GGA functionals (ωB97,[37] ωB97X,[37] and ωB97X-D[38]), and two global hybrid meta-GGA functionals (M06-HF[36, 39] and M06-2X[40]).

All calculations are performed using the Gaussian09 program.[41] Results are computed using the aug-cc-pVQZ basis set with the ultrafine grid, EML(99,590), consisting of 99

Euler-Maclaurin radial grid points[42] and 590 Lebedev angular grid points.[43] All the calculated results are provided in the Supporting Material (Table S1 to S12). The error for each entry is defined as error = theoretical value − reference value. The notations used for characterizing statistical errors are as follows: mean signed errors (MSEs), mean absolute errors (MAEs), and root-mean-square (RMS) errors.

### A. Vertical ionization potentials

The vertical ionization potential (IP) of a neutral molecule is defined as the energy difference between the cationic and neutral charge states,

$$IP(1) = E_{tot} (cation) - E_{tot} (neutral). \qquad (1)$$

For the exact KS-DFT, the vertical IP of a neutral molecule is the same as the minus HOMO (highest occupied molecular orbital) energy of the neutral molecule,[32, 44]

$$IP(2) = -E_{HOMO} (neutral) \qquad (2)$$

Therefore, IP(2) is the same as IP(1) for the exact KS-DFT. For approximate density functional methods, IP(1) and IP(2) may be adopted to examine the accuracy of the predicted total energies and HOMO energies, respectively.

### B. Vertical electron affinities

The vertical electron affinity (EA) of a neutral molecule is defined as the energy difference between the neutral and anionic charge states,

$$EA(1) = E_{tot} \text{ (neutral)} - E_{tot} \text{ (anion)}. \tag{3}$$

By comparing Eq. (1) with Eq. (3), the vertical EA of a neutral molecule is identical to the vertical IP of the anion, which is, for the exact KS-DFT, the minus HOMO energy of the anion,

$$EA(2) = -E_{HOMO} \text{ (anion)}. \tag{4}$$

For the exact KS-DFT, EA(2) is identical to EA(1). Therefore, EA(1) and EA(2) may be adopted to examine the accuracy of approximate density functional methods on total energies and HOMO energies, respectively.

In addition, the vertical EA of a neutral molecule is conventionally approximated by the minus LUMO (lowest unoccupied molecular orbital) energy of the neutral molecule,[18]

$$EA(3) = -E_{LUMO} \text{ (neutral)}. \tag{5}$$

However, even for the exact KS-DFT, a difference exists between EA(3) and vertical EA due to the derivative discontinuity (DD) of the exchange-correlation functional.[45-48] Recent study shows that DD is close to zero for LC hybrid functionals,[49] so the EA(3) calculated by a LC hybrid functional should be close to the true vertical EA.

**C. Fundamental gaps**

The fundamental gap (Eg) of a molecule is defined as the difference between the vertical

IP and vertical EA of the molecule,

$$Eg = IP - EA. \tag{6}$$

Since there are many different ways of calculating vertical IP and vertical EA from density functional methods, we have mainly three ways of calculating Eg:

$$Eg(1) = IP(1) - EA(1) = E_{tot}\,(\text{cation}) + E_{tot}\,(\text{anion}) - 2\,E_{tot}\,(\text{neutral}) \tag{7}$$

$$Eg(2) = IP(2) - EA(2) = E_{HOMO}\,(\text{anion}) - E_{HOMO}\,(\text{neutral}) \tag{8}$$

$$Eg(3) = IP(2) - EA(3) = E_{LUMO}\,(\text{neutral}) - E_{HOMO}\,(\text{neutral}) \tag{9}$$

Note that Eg(3) is the so-called Kohn-Sham (KS) gap or HOMO-LUMO gap in KS-DFT.[18, 50] As mentioned previously, for LC hybrid functionals, EA(3) should be close to vertical EA, and hence Eg(3) should be close to the true Eg.

**D. Optical gaps**

The optical gap (Eopt) of a molecule, defined by a neutral excitation, is the energy difference between the lowest dipole-allowed excited state and the ground state,

$$E_{opt} = E_{tot}\,(\text{neutral}) - E_{tot}\,(\text{neutral}). \tag{10}$$

As Eopt is an excited-state property, it cannot be directly calculated using CCSD or KS-DFT. To be consistent with the ground-state calculations, we adopt EOM-CCSD[22-30] and TDDFT[51-57] (with the same density functionals for the ground-state calculations) to calculate the Eopt.

**E. Exciton binding energies**

The exciton binding energy (Eb) of a molecule is defined as the difference between the fundamental and optical gaps,

$$Eb = Eg - Eopt. \qquad (11)$$

As there are three ways of calculating Eg (see Eq. (7) – Eq. (9)), Eb can also be calculated in three ways, i.e.,

$$Eb(1) = Eg(1) - Eopt \qquad (12)$$

$$Eb(2) = Eg(2) - Eopt \qquad (13)$$

$$Eb(3) = Eg(3) - Eopt \qquad (14)$$

**III. Results and discussion**

**A. Benchmark methods**

The IP(1) calculated from CCSD and CCSD(T) are compared to experimental values for the 121 molecules (see Figure S1 and Table S13 in Supplemental Material). Both methods are very accurate with MAE being 0.17 eV from CCSD and 0.14 eV from CCSD(T). The difference between these two methods are around 0.1 eV for IP(1) and Eg(1), and around 0.05 eV for EA(1) (see Figure S1 and Table S14 Supplemental Material). Such differences are much smaller than those associated with density functional methods (usually greater than 0.5 eV). Therefore, we believe that the CCSD results can be taken as benchmarks. In this study,

CCSD and EOM-CCSD (i.e., for consistency with CCSD) are adopted as the benchmarking methods for the ground-state and excited-state properties, respectively.

**B. Accuracy of density functional methods**

Five optoelectronic properties, including the vertical ionization potentials, vertical electron affinities, fundamental gaps, optical gaps, and exciton binding energies of 121 molecules are adopted to examine the accuracy of various density functionals in KS-DFT and TDDFT.

The accuracy of various density functionals with respect to experimental values on IP(1) and IP(2) are summarized in Table S15 in Supplemental Material. Despite the fact that different measures of accuracy yield slightly different ranks in performance of IP(1), M06-2X and ωB97X-D are among the most accurate density functionals, with an MAE of around 0.18 eV. LDA is the least accurate with an MAE of 0.56 eV, which is more than three times the error of the best functional. It is also interesting to note that while M06-HF and M06-2X are both hybrid meta-GGA functionals, the MAE of M06-HF (0.46 eV) is more than twice that of M06-2X (0.18 eV).

There is a significant variation in the performance when the ionization potentials are estimated from the HOMO energy (IP(2)). The ωB97 functional is the best functional with an MAE of 0.42 eV. PBE is the least accurate with an MAE of 4.44 eV, which is more ten times higher than that of the best functional here. The MAE difference in IP(1) between M06-HF and M06-2X is more than twice, but the performance of M06-HF (0.99 eV) and M06-2X (1.51 eV) is similar for IP(2). In general, the MAE in IP(2) is 2 to 13 times larger than those for IP(1).

Since there are no comprehensive experimental data available for the other properties, the results from CCSD are taken as the reference values for evaluating the performance of density functional methods. Table 1 shows such comparison for the vertical ionization potentials, vertical electron affinities, fundamental gaps, optical gaps, and exciton binding energies. Figures 1 to 5 illustrates the MAE in IP, EA, Eg, Eopt, and Eb with different definitions. For IP(1), M06-2X agrees best with CCSD with an MAE difference of 0.13 eV. The ωB97 series also provide quite consistent results with CCSD with MAE difference being about 0.16 eV. For EA(1), the performance of M06-2X and the ωB97 series are among the best, with the ωB97X-D being the best functional (MAE = 0.16 eV). The ωB97X and M06-2X show the best performance for Eg(1) are shown that ωB97X (MAE = 0.26 eV) and M06-2X (MAE = 0.27 eV) are the best functionals. The LDA yields the worst results for IP(1) and EA(1); however, the PBE is the least accurate for Eg(1) because it generally underestimates IP(1) (MAE=0.42 eV, MSE=-0.31 eV) but overestimates EA(1) (MAE=0.34 eV, MSE=0.32 eV). In summary, M06-2X and the ωB97 series (ωB97, ωB97X, and ωB97X-D) are the best density functionals for the ground-state properties (with an accuracy of 0.3 eV or less). Furthermore, as in the case of IP, the MAE from the least accurate methods (often LDA and GGAs) can be more than twice that of the best functional.

The optical gap is a measure of performance in describing the excited state. The ωB97 and ωB97X functionals perform the best for optical gaps. PBE shows surprisingly poor results for Eopt, with an MAE (1.12 eV) that is three times higher than the best method.

Since ωB97 and ωB97X perform the best for both Eg(1) and Eopt, unsurprisingly, they are the most accurate methods for Eb(1). ωB97X-D has 0.1 eV more than ωB97 and ωB97X for Eopt, but ωB97X-D also has good performance as ωB97 and ωB97X. The MAE in Eb(1) from M06-HF (0.80 eV) is more than twice larger than that from ωB97X. It is interesting to note that while B3LYP is inaccurate for Eg(1) (MAE=0.45 eV) and Eopt (MAE=0.76 eV), it is reasonably accurate for Eb(1) (MAE=0.53 eV), which is a result of systematic underestimation (IP(1): MAE=0.26 eV, MSE=0.09 eV), EA(1): MAE=0.31 eV, MSE=0.30 eV), Eg(1) (MAE=0.45 eV, MSE=-0.39 eV), and Eopt (MAE=0.76 eV, MSE=-0.63 eV). In summary, the ωB97 functional provides the most reliable predictions for total energy of a molecule both in the ground state and in the excited state.

The performance in IP(2), EA(2), Eg(2), and Eb(2) is an indication of the quality of HOMO energy. In general, the MAE for IP(2) is 3 to 9 times larger than those for IP(1). ωB97 is the best functional for IP(2) with an MAE of 0.41 eV (compared to 0.16 for IP(1)). The ωB97X-D underestimated the HOMO energy, resulting in a large error in IP(2) (MAE = 1.13 eV and MSE = -1.12 eV), Eg(2) and Eb(2). The MAE errors from ωB97 methods for EA(2) are quite similar to those for EA(1). However, LDA, PBE, M06L, and B3LYP show significantly increased inaccuracy in EA(2). As a result, these methods are also poor for Eg(2) and Eb(2). The ωB97 is the best functional for Eg(2) (MAE = 0.58 eV) and Eb(2) (MAE = 0.53 eV). The global hybrid MGGA functionals, M06-2X and M06-HF, also provide reasonable accuracy in the HOMO energies. However, since M06-2X underestimates for Eopt and Eg, it is accurate for Eb(2) which is also a result of systematic underestimation.

The performance in EA(3), Eg(3), and Eb(3) implies the description accuracy in the LUMO energy. The performance of ωB97 series in these properties are quite similar to the corresponding EA(2), Eg(2), and Eb(2). However, all other methods, including hybrid MGGA methods, show significantly increased MAEs here. For example, the MAE of EA(3) from M06-HF (1.80 eV) is almost 3 times higher than that of EA(2) (0.55 eV). Therefore, the ωB97 series provide reliable description for the LUMO energy.

It is noteworthy that in some cases the calculated LUMO energies are so small that Eg(3) becomes smaller than optical gap, yielding qualitatively incorrect exciton binding energies (i.e., Eb can be negative!). 115 out of 121 from LDA, 116 out of 121 from PBE, 112 out of 121 from M06L, 38 out of 121 from B3LYP, 10 out of 121 from M06-2X and 7 out of 121 from M06-HF, 2 out of 121 from ωB97 and ωB97X, 5 out of 121 from ωB97X-D show negative Eb(3) (see Table S12 for a complete list of all the Eb(3).) In some cases, the calculated LUMO energies are even lower than the calculated HOMO energies from M06 methods (M06L, M06-HF, M06-2X), resulting in incorrect negative Eg(3) (-87.12 eV (Hydrogen atom) from M06L, -5.00 eV (Hydrogen atom) from M06-2X and -40.36 eV (Hydrogen atom), -18.76 eV (Lithium atom), -13.04 eV (Sodium atom) from M06-HF).

**IV. Conclusions**

In conclusion, we have examined the performance of several density functional methods on the exciton binding energies and related optoelectronic properties of 121 small- to medium-sized molecules. Relative to the highly accurate CCSD and EOM-CCSD methods, ωB97, ωB97X, and ωB97X-D exhibit the best accuracy for these properties. However, when intermolecular charge-transfer excitons are involved, ωB97X-D, which includes dispersion corrections, is expected to be essential.

**Acknowledgements**

We thank the support from the Ministry of Science and Technology of Taiwan (Grant Nos.: NSC101-2628-E-002-014-MY3 and MOST104-2628-M-002-011-MY3), National Taiwan University (Grant Nos.: NTU-CDP-104R7876 and NTU-CDP-104R7818), and the National Center for Theoretical Sciences of Taiwan. The computational resources from the National Center for High-Performance Computing of Taiwan and the Computing and Information

Networking Center of the National Taiwan University are acknowledged.

**Table 1. Statistical errors (in eV) of 9 density functional methods for various properties (with respect to the CCSD or EOM-CCSD data).**

|        | Error | LDA   | PBE   | M06L  | B3LYP | ωB97  | ωB97X | ωB97X-D | M06-2X | M06-HF |
|--------|-------|-------|-------|-------|-------|-------|-------|---------|--------|--------|
| IP(1)  | MSE   | 0.39  | -0.31 | -0.36 | -0.09 | -0.07 | -0.07 | -0.09   | **0.03**  | 0.35   |
|        | MAE   | 0.58  | 0.42  | 0.39  | 0.26  | 0.16  | 0.17  | 0.20    | **0.13**  | 0.37   |
|        | RMS   | 0.66  | 0.61  | 0.56  | 0.37  | 0.23  | 0.24  | 0.28    | **0.18**  | 0.46   |
| IP(2)  | MSE   | -3.92 | -4.50 | -4.35 | -3.25 | **-0.34** | -0.59 | -1.12   | -1.57  | 0.84   |
|        | MAE   | 3.92  | 4.50  | 4.35  | 3.25  | **0.41**  | 0.59  | 1.13    | 1.57   | 0.91   |
|        | RMS   | 4.05  | 4.61  | 4.45  | 3.32  | **0.67**  | 0.82  | 1.28    | 1.63   | 1.04   |
| EA(1)  | MSE   | 0.78  | 0.32  | -0.07 | 0.30  | **0.02**  | 0.07  | 0.13    | 0.05   | 0.17   |
|        | MAE   | 0.79  | 0.34  | 0.24  | 0.31  | 0.20  | 0.18  | **0.16**    | 0.21   | 0.33   |
|        | RMS   | 0.86  | 0.46  | 0.38  | 0.43  | 0.34  | 0.32  | **0.31**    | 0.37   | 0.42   |
| EA(2)  | MSE   | -1.27 | -1.62 | -1.92 | -1.22 | 0.13  | 0.12  | **0.00**    | -0.75  | 0.34   |
|        | MAE   | 1.31  | 1.67  | 1.95  | 1.28  | 0.24  | **0.20**  | 0.24    | 0.80   | 0.55   |
|        | RMS   | 1.60  | 1.97  | 2.20  | 1.47  | 0.39  | **0.35**  | 0.39    | 0.87   | 0.76   |
| EA(3)  | MSE   | 3.15  | 2.53  | 2.87  | 1.95  | -0.07 | **-0.03** | 0.24    | 1.25   | 1.75   |
|        | MAE   | 3.15  | 2.53  | 2.87  | 1.96  | **0.29**  | 0.34  | 0.42    | 1.26   | 1.80   |
|        | RMS   | 3.44  | 2.81  | 8.90  | 2.15  | **0.37**  | 0.41  | 0.56    | 1.80   | 5.57   |
| Eg(1)  | MSE   | -0.39 | -0.63 | -0.29 | -0.39 | -0.09 | -0.14 | -0.22   | **-0.02** | 0.18   |
|        | MAE   | 0.51  | 0.65  | 0.44  | 0.45  | 0.28  | **0.26**  | 0.30    | 0.27   | 0.33   |
|        | RMS   | 0.71  | 0.88  | 0.62  | 0.62  | 0.42  | 0.41  | 0.44    | **0.40**   | 0.46   |
| Eg(2)  | MSE   | -5.46 | -2.88 | -2.43 | -2.03 | **-0.47** | -0.71 | -1.13   | -0.82  | 0.49   |
|        | MAE   | 6.14  | 2.88  | 2.43  | 2.03  | **0.58**  | 0.73  | 1.13    | 0.86   | 0.75   |
|        | RMS   | 8.08  | 3.14  | 2.70  | 2.22  | **0.82**  | 0.94  | 1.30    | 1.01   | 0.95   |
| Eg(3)  | MSE   | -7.07 | -7.03 | -7.22 | -5.19 | **-0.27** | -0.56 | -1.36   | -2.82  | -0.92  |
|        | MAE   | 7.07  | 7.03  | 7.22  | 5.19  | **0.51**  | 0.64  | 1.39    | 2.82   | 1.57   |
|        | RMS   | 7.32  | 7.25  | 11.25 | 5.35  | **0.77**  | 0.93  | 1.61    | 3.21   | 5.66   |
| Eopt | MSE | -0.91 | -1.05 | -0.61 | -0.63 | **-0.13** | -0.15 | -0.33 | -0.38 | -0.44 |
| | MAE | 0.97 | 1.12 | 0.70 | 0.76 | **0.31** | 0.32 | 0.46 | 0.46 | 0.68 |
| | RMS | 1.15 | 1.29 | 0.90 | 0.92 | 0.55 | 0.63 | 0.71 | **0.54** | 1.05 |
| Eb(1) | MSE | 0.52 | 0.42 | 0.32 | 0.24 | 0.04 | **0.01** | 0.12 | 0.36 | 0.62 |
| | MAE | 0.66 | 0.66 | 0.57 | 0.53 | 0.39 | **0.38** | 0.40 | 0.48 | 0.80 |
| | RMS | 0.88 | 0.88 | 0.74 | 0.78 | 0.71 | 0.71 | 0.71 | **0.61** | 1.16 |
| Eb(2) | MSE | -1.74 | -1.83 | -1.82 | -1.39 | **-0.34** | -0.56 | -0.79 | -0.43 | 0.93 |
| | MAE | 1.77 | 1.85 | 1.86 | 1.42 | **0.53** | 0.66 | 0.87 | 0.57 | 1.17 |
| | RMS | 2.05 | 2.13 | 2.12 | 1.66 | 0.81 | 0.92 | 1.11 | **0.71** | 1.49 |
| Eb(3) | MSE | -6.16 | -5.98 | -6.62 | -4.56 | **-0.13** | -0.41 | -1.03 | -2.44 | -0.48 |
| | MAE | 6.16 | 5.98 | 6.62 | 4.56 | **0.48** | 0.59 | 1.11 | 2.44 | 1.69 |
| | RMS | 6.48 | 6.28 | 10.91 | 4.80 | **0.74** | 0.96 | 1.46 | 2.86 | 5.69 |
| Total | MSE | -1.84 | -1.88 | -1.87 | -1.36 | **-0.14** | -0.25 | -0.47 | -0.63 | 0.32 |
| | MAE | 2.75 | 2.47 | 2.47 | 1.83 | **0.36** | 0.42 | 0.65 | 0.99 | 0.91 |
| | RMS | 4.08 | 3.45 | 5.50 | 2.57 | **0.60** | 0.69 | 0.96 | 1.52 | 2.93 |

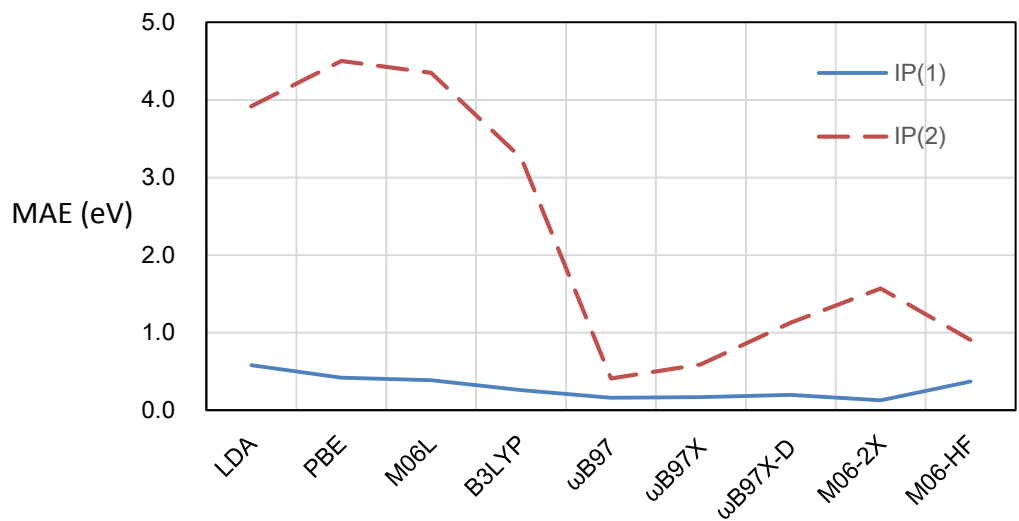

Figure 1. Comparison of MAE in ionization potential (IP) from 9 DFT methods with respective to CCSD. IP(1) (blue solid line) is calculated from eq. 1 and IP(2) (red dashed line) from eq. 2.

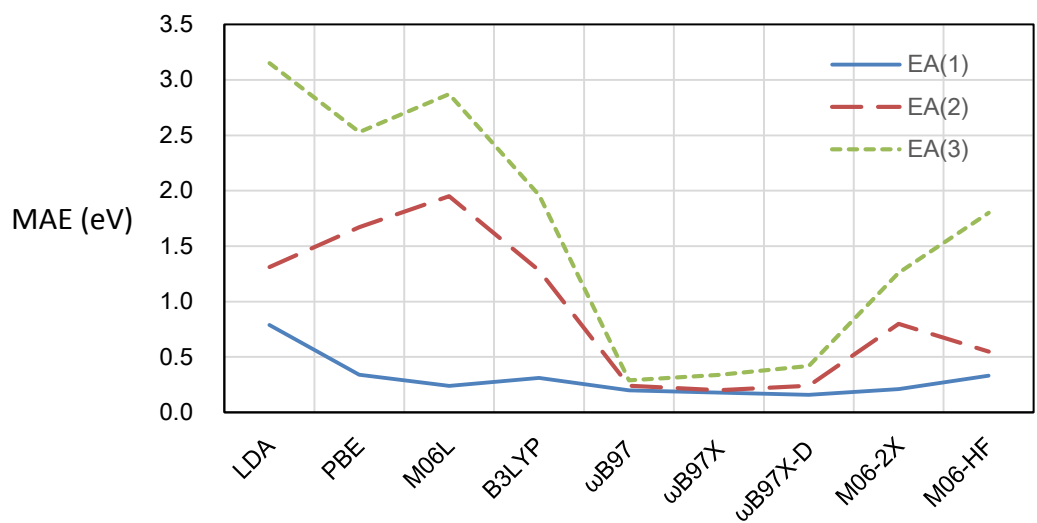

Figure 2. Comparison of MAE in electron affinity (EA) from 9 DFT methods with respective to CCSD. EA(1) (blue solid line) is calculated from eq. 3, EA(2) (red dashed line) from eq. 4, and EA(3) from eq. 5.

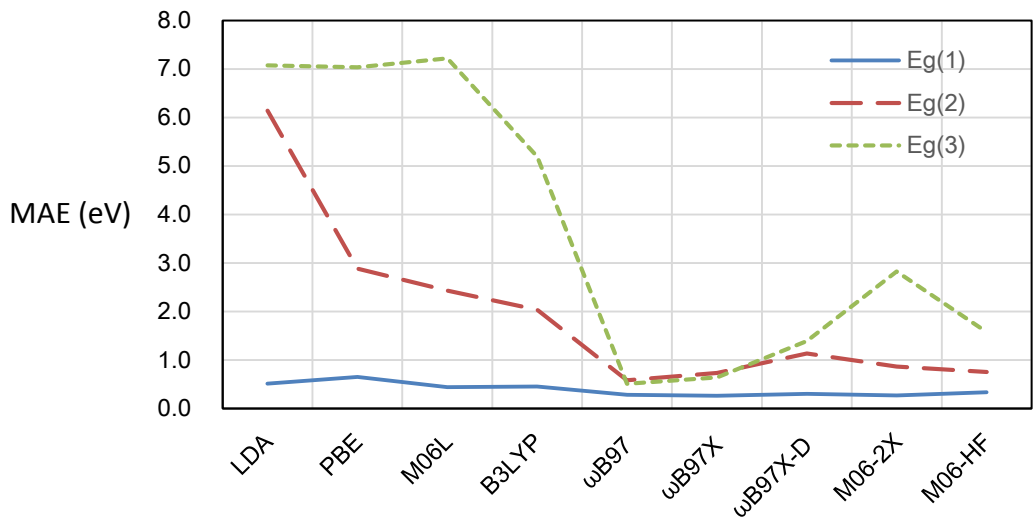

Figure 3. Comparison of MAE in fundamental gap (Eg) from 9 DFT methods with respective to CCSD. Eg(1) (blue solid line) is calculated from eq. 7, Eg(2) (red dashed line) from eq. 8, and Eg(3) from eq. 9.

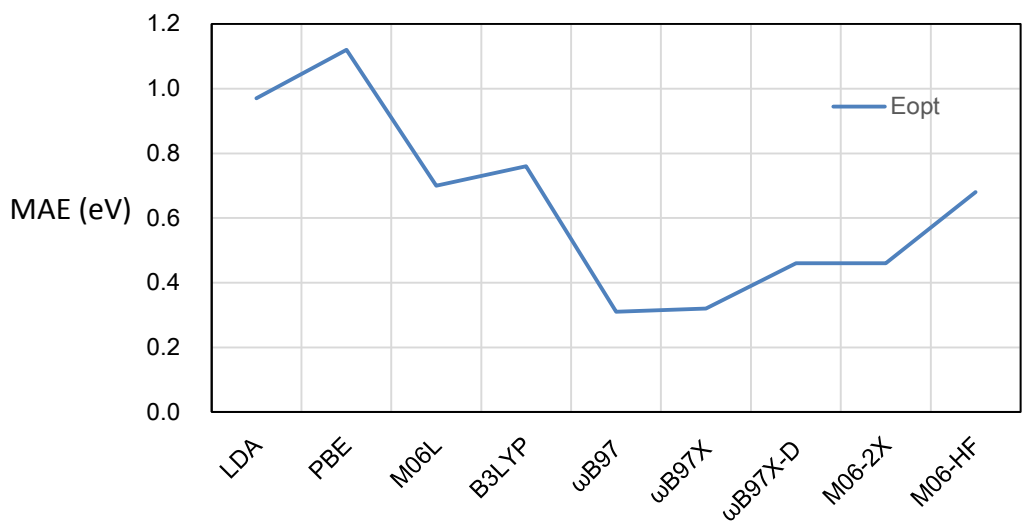

Figure 4. Comparison of MAE in optical gap (Eopt from eq. 10) from 9 DFT methods with respective to EOM-CCSD.

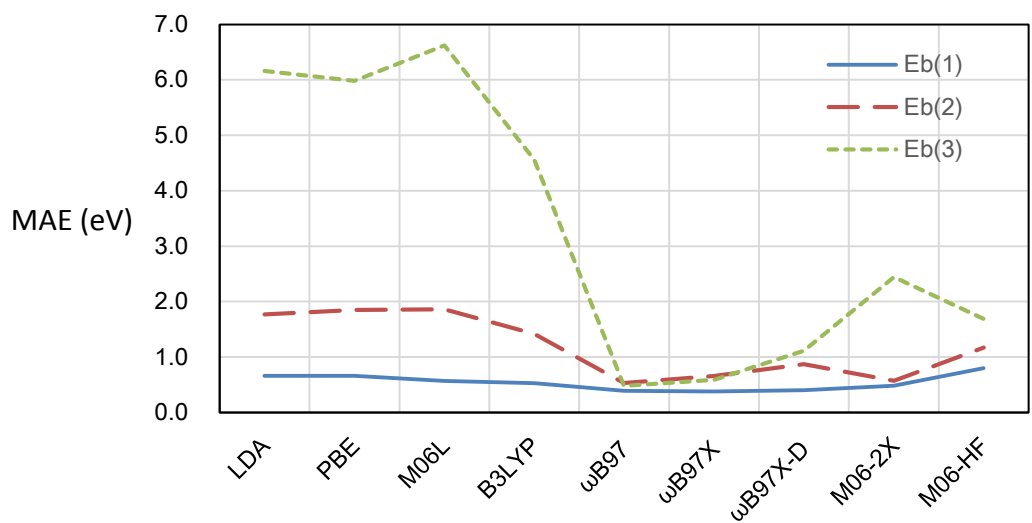

Figure 5. Comparison of MAE in binding energy (Eb) from 9 DFT methods with respective to CCSD and EOM-CCSD. Eb(1) (blue solid line) is calculated from eq. 12, Eb(2) (red dashed line) from eq. 13, and Eb(3) from eq. 14.